\documentclass[twocolumn,aps,prl,tightenlines]{revtex4-2}%superscriptaddress,
\usepackage{amsmath,amsthm,amssymb}
\usepackage{latexsym}
\usepackage{amscd,graphicx}
\usepackage{epsfig}
\usepackage[normalem]{ulem}
\usepackage[dvipsnames]{xcolor}
\usepackage{booktabs}
\graphicspath{{figs/}}
\usepackage{placeins}
\usepackage{float}
\usepackage[colorlinks=true, linkcolor=red, citecolor=blue, urlcolor=blue,
            filecolor=blue, breaklinks=true, pdfborder={0 0 0}]{hyperref}
\makeatletter
\def\@textbottom{\vskip 0pt}
\let\@texttop\relax
\makeatother

\renewcommand{\Im}{\mathrm{Im}}
\renewcommand{\Re}{\mathrm{Re}}

\begin{document}

\title{Quantum Battery Diode}

\author{Zi-Jie Wang}
\affiliation{Department of Physics, Wenzhou University, Zhejiang 325035, China}

\author{Jiayu Gao}
\affiliation{Department of Physics, Wenzhou University, Zhejiang 325035, China}

\author{Sheng-Bo Zheng}
\affiliation{Department of Physics, Wenzhou University, Zhejiang 325035, China}

\author{Jiaojiao Chen}
\altaffiliation{jjchenphys@wzu.edu.cn}
\affiliation{Department of Physics, Wenzhou University, Zhejiang 325035, China}

\author{Wei Xiong}
\altaffiliation{xiongweiphys@wzu.edu.cn}
\affiliation{Department of Physics, Wenzhou University, Zhejiang 325035, China}

\date{\today}

\begin{abstract}
Suppressing energy backflow is crucial for practical quantum batteries (QBs). Here, we propose a QB consisting of three bosonic modes arranged in a closed loop threaded by a synthetic gauge flux to mitigate this unwanted effect. We first show that the direction of energy flow in the loop can be flexibly controlled by the introduced synthetic flux, resulting in nonreciprocal enhancement of the QB performance. Second, we numerically demonstrate that the energy backflow from the charger to the battery can be fully suppressed at quarter flux in the transient regime, as quantified by the defined time-dependent net energy current. Third, we show that the reverse coupling (battery to charger) can be fully switched off when the flux-specified linewidth-matching condition, i.e., at the diode point, is satisfied. At this point, the forward coupling (charger to battery) is doubled, giving rise to significant improvement of the steady-state storage and peak efficiency of the battery. Finally, the effects of fluctuations, thermal noise, and dephasing on isolation, as well as the experimental realization, are provided. Our work offers a potential route to perfectly suppress energy backflow in coherently coupled quantum batteries, thereby advancing their practical application.
\end{abstract}

\maketitle

{\it Introduction}.---Quantum batteries (QBs), first proposed by Alicki and Fannes in $2013$~\cite{Alicki2013}, have attracted considerable interest in quantum information science and thermodynamics~\cite{Campaioli2024}. Quantum batteries can exploit quantum features for energy storage and release and can potentially outperform classical counterparts, with enhanced charging power~\cite{Campaioli2017,Binder2015,Francica2020,Shi2022,Seah2021,Zhu2023,Centrone2023,Rossini2020,Gyhm2023,Gyhm2022,Rosa2020,Rodriguez2023,Mazzoncini2023,Konar2024b,Zhang2019}, increased capacity~\cite{Yang2023cap,JuliaFarre2020,Tirone2024,Yang2024exp,Gao2022,Zhang2024}, and superior work extraction~\cite{Allahverdyan2004,Andolina2019,Hovhannisyan2013,GarciaPintos2020,Wang2021fluct,Song2024entropic,Castellano2025,Francica2024,Monsel2020,Tirone2023,Tirone2025}. In the past decade, a variety of QB models have been proposed theoretically, including the Dicke type~\cite{Ferraro2018,Dou2022}, spin-chain type~\cite{Le2018,Rossini2020}, oscillator type~\cite{Andolina2025,Downing2025}, etc.~\cite{Yang2024,YangH2025,Liu2021central,Canzio2025,Hao2025,Ali2024,Lu2025,Ezawa2026,Friis2018,Konar2024,Bai2020,Song2024,Song2025,YangAn2026,Yan2026}.  Experimentally, QBs have been demonstrated with NMR spins~\cite{Joshi2022}, superconducting circuits~\cite{Hu2022,Massa2025,Elghaayda2025,Li2025ergo,Hu2026}, organic microcavities~\cite{Quach2022}, and photonic platforms~\cite{Zhu2023,Huang2023}. However, these QBs are mainly built on coherent charger-battery coupling. This mechanism allows the stored energy in the battery to flow back to the charger, namely energy backflow, which severely limits the application of QBs. Therefore, suppressing energy backflow has become an essential challenge in practical QBs. 

To address this challenge, various strategies have been proposed. Early approaches employed switchable charger-battery couplings to prevent energy from flowing back to the charger~\cite{Farina2019,Santos2019}. Subsequently, specially engineered quantum states and transitionless quantum driving were used in place of switchable couplings~\cite{Santos2020,Moraes2022}. More recently, optimal control~\cite{Mazzoncini2023,Lai2024}, dynamical modulation~\cite{Hadipour2025}, quantum Zeno control~\cite{Xie2026}, and engineered reservoirs and dissipative mediators~\cite{Ahmadi2025,Kamin2024,ZhangJing2025,Song2024} have been explored to regulate energy transfer and enhance energy retention. In particular, reservoir engineering has enabled nonreciprocal energy transfer, providing a route to suppress backflow without simply interrupting the charger-battery coupling~\cite{Metelmann2015,Metelmann2017,Ranzani2015,Ahmadi2024,Zafar2026}. However, such a method requires carefully balancing the coherent and dissipative couplings induced by the common reservoir.

In this Letter, we propose an alternative approach to suppress energy backflow by introducing a synthetic gauge-invariant phase into a ring bosonic QB comprising a charger coherently coupled to two linked batteries. We first show that the first-moment dynamics of the QB is confined to an invariant real manifold at quarter flux, ensuring a nonnegative net energy current into the storage battery over a finite damping range. This indicates that the backflow can be dynamically suppressed. Then, fixing the quarter flux and achieving the linewidth-matching condition, equivalently, operating the QB at the diode point, we find that the reverse coupling (battery to charger) can be fully switched off, while the forward coupling (charger to battery) is doubled. As a result, the steady-state storage and the peak efficiency of the battery at this point are greatly enhanced. By studying the parameter mismatch, we find that the reverse isolation is strongly robust against parameter mismatch. Specifically, a $1\%$ fluctuation leads to $\sim46$ dB isolation. Finally, we show that thermal noise limits the total-energy isolation, which scales as ($1/n_{\mathrm{th}}$), while dephasing shifts the linewidth-matching condition. This shift can be compensated for by retuning the linewidth of the assisted battery. These results establish a potential route to perfectly suppressing energy backflow in coherently coupled QB devices.

\begin{figure}
	\centering
	\includegraphics[width=0.42\textwidth]{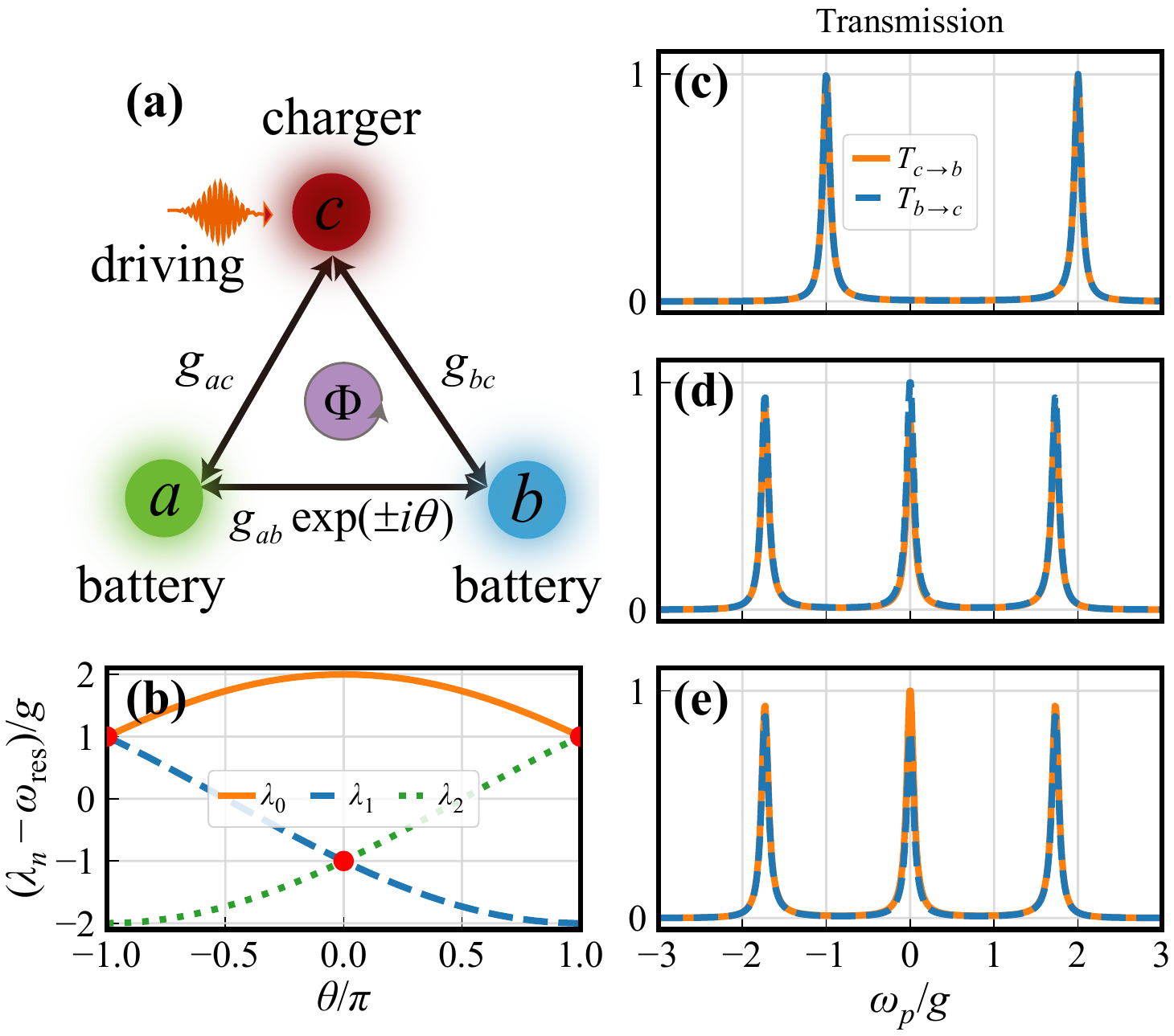}
	\caption{(a) Ring quantum battery with charger $c$ and cells $a$, $b$ sharing a loop phase $\theta$. (b) Spectrum $(\lambda_n-\omega_{\rm res})/g$ of the undriven ring vs $\theta/\pi$, red circles marking twofold degeneracies. (c)-(e) Transmission $T_{c\rightarrow b}$ and $T_{b\rightarrow c}$ vs $\omega_p/g$ for (c) $\theta=0$, (d) $\theta=\pi/2$, (e) $\theta=-\pi/2$. Here, $g_{ac}=g_{bc}=g_{ab}=g$, $\kappa_\alpha=\kappa=0.1g$.}
	\label{figure1}
\end{figure}

{\it Model and Hamiltonian}.---As illustrated in Fig.~\ref{figure1}(a), we consider a flux-threaded bosonic ring consisting of a driven resonator $c$ as the charger and two undriven resonators $a$ and $b$ as the battery cells. Within the rotating-wave
approximation, the Hamiltonian reads (setting $\hbar=1$)~\cite{SM}
\begin{align}
	H=H_{\rm driven}+\big(
	g_{ac}a^\dagger c
	+g_{bc} b c^\dagger
	+g_{ab}e^{i\theta}a^\dagger b
	+{\rm H.c.}
	\big),
	\label{e1}
\end{align}
where $H_{\rm driven}=\omega_{\rm res}\sum_\alpha\alpha^\dagger\alpha+H_d$, with $\omega_{\rm res}$ the resonant frequency, $\alpha=a,b,c$, and 
$H_d=i\Omega_d (c^\dagger e^{-i\omega_d t}-c\,e^{i\omega_d t})$.
Here, $\omega_d$ ($\Omega_d$) is the frequency (amplitude) of the driving field, $g_{ac}$, $g_{bc}$, and $g_{ab}$ are real coupling strengths, while
$\theta$ is the gauge-invariant phase accumulated around the ring,
determined by the magnetic flux
$\Phi=(\theta/2\pi)\Phi_0$, where $\Phi_0$ is the flux quantum~\cite{Peierls1933,Dalibard2011}. In the absence of driving ($\Omega_d=0$), the Hamiltonian in Eq.~(\ref{e1}) in the uniform-coupling limit (i.e., $g_{ac}=g_{bc}=g_{ab}\equiv g$) can be diagonalized as $H=\sum_n\lambda_n d_n^\dagger d_n$, where
$d_n=(c+e^{i\phi_n}a+e^{-i\phi_n}b)/\sqrt{3}$,
$\phi_n=(\theta+2n\pi)/3$, and
$\lambda_n=\omega_{\rm res}+2g\cos\phi_n$.
The spectrum is periodic in $\theta$ and invariant under
$\theta\rightarrow-\theta$, as depicted in Fig.~\ref{figure1}(b). Two eigenvalue branches become degenerate at
$\theta=m\pi$, where time-reversal symmetry is preserved. Away from
these points, a nontrivial flux lifts the degeneracy and distinguishes
the two orientations of the ring, with the spectral ordering reversed
upon changing the sign of $\theta$.
Thus, although the spectrum is invariant under flux reversal, the corresponding eigenmodes retain information about the orientation of the loop.

The flux also induces nonreciprocal transmission between the
charger and the battery cells~\cite{SM}. At $\theta=m\pi$,
time-reversal symmetry ensures reciprocal transmission,
$T_{c\to b}=T_{b\to c}$ [Fig.~\ref{figure1}(c)]. Away from
these points, the two transmission directions become
inequivalent, and reversing the flux interchanges their
responses. The two transmission coefficients vs $\omega_p/g$ at the quarter-flux point with $\kappa=0.1g$ are shown in Figs.~\ref{figure1}(d) and \ref{figure1}(e). Specifically, $T_{c\to b}=0.400$ and
$T_{b\to c}=0.489$ at $\theta=+\pi/2$
[Fig.~\ref{figure1}(d)], with the two values interchanged
at $\theta=-\pi/2$ [Fig.~\ref{figure1}(e)].
These results demonstrate that the flux controls the direction
of transmission, while the degree of nonreciprocity depends
on the system parameters~\cite{SM}.

\begin{figure}
	\centering
	\includegraphics[width=0.42\textwidth]{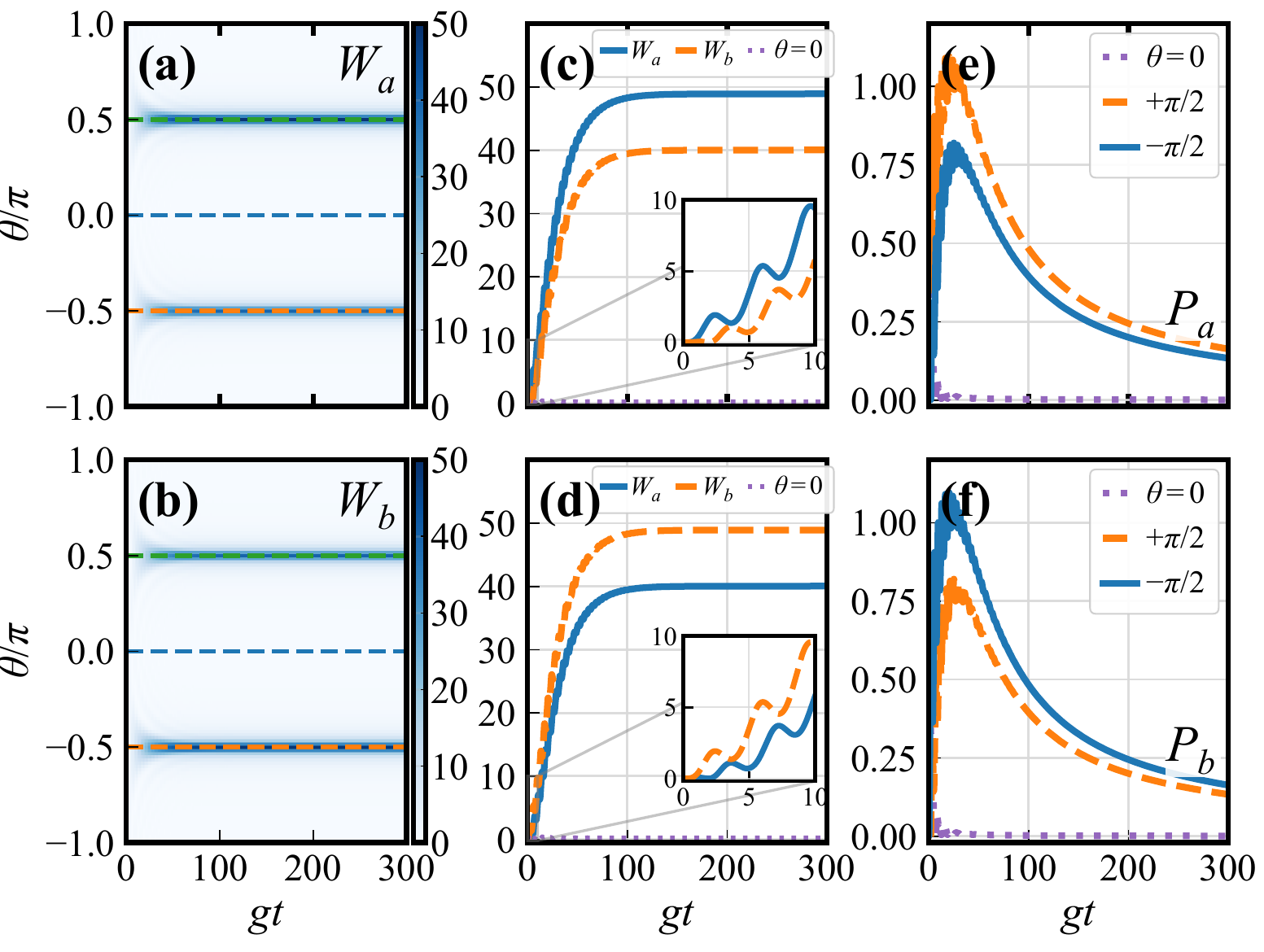}
	\caption{(a) Ergotropy $W_a$ and (b) $W_b$ vs $gt$ and $\theta/\pi$, with dashed lines at $\theta/\pi=0.5$ (green), $0$ (blue), $-0.5$ (orange). (c) $W_a$ and (d) $W_b$ vs $gt$ at (c) $\theta=+\pi/2$ and (d) $\theta=-\pi/2$, with dotted curves for the $\theta=0$ reference of cell $b$ and insets for short-time dynamics. (e) Power $P_a$ and (f) $P_b$ vs $gt$ for $\theta=0$ (purple dotted), $+\pi/2$ (orange dashed), $-\pi/2$ (blue solid). Here, $\kappa=0.1g$, $g=\Omega_d=1$, $\Delta=0$.}
	\label{fig2}
\end{figure}

{\it Nonreciprocal charging}.---
To quantify the charging dynamics, we now include dissipation~\cite{Lindblad1976}. The first moments, obtained from the input-output formalism~\cite{GardinerCollett1985}, obey
$\dot{\mathbf M}=-\mathbf A\mathbf M+\mathbf D$, with
$\mathbf M=[\langle c\rangle,\langle a\rangle,\langle b\rangle]^T$,
$\mathbf D=[\Omega_d,0,0]^T$, and
\begin{align}
	\mathbf A=
	\begin{pmatrix}
		\kappa_c/2+i\Delta & ig_{ac} & ig_{bc}\\
		ig_{ac} & \kappa_a/2+i\Delta & ig_{ab}e^{i\theta}\\
		ig_{bc} & ig_{ab}e^{-i\theta} & \kappa_b/2+i\Delta
	\end{pmatrix}.
	\label{eqAmat}
\end{align}
The formal solution is
$\mathbf M(t)=\mathbf M_{\rm ss}+e^{-\mathbf A t}
[\mathbf M(0)-\mathbf M_{\rm ss}]$, with
$\mathbf M_{\rm ss}=\mathbf A^{-1}\mathbf D$.
Starting from vacuum,
$\mathbf M(t)=(\mathbf 1-e^{-\mathbf A t})\mathbf M_{\rm ss}$,
and $N_{a(b)}(t)=|[\mathbf M(t)]_{a(b)}|^2$.
For the linear, number-conserving dynamics with vacuum input, an
initial vacuum evolves into a product of coherent states. Thus, at zero
temperature, $W_{a(b)}=\omega_{\rm res} N_{a(b)}$, and the average charging
power is $P_{a(b)}=W_{a(b)}/t$.  In Figs.~\ref{fig2}(a) and \ref{fig2}(b), the two ergotropies as functions of $gt$ and $\theta/\pi$ are plotted. At fixed flux, both grow with damped
oscillations before approaching finite steady values
$W_{a(b)}(\infty)=\omega_{\rm res} N_{a(b)}(\infty)$, where $N_{a(b)}(\infty)
={(g\Omega_d)^2
	\left(g^2+\kappa^2/4\pm g\kappa\sin\theta\right)}f^{-1}(\theta)
$ with $f(\theta)=
\left(\kappa^3/8+3g^2\kappa/2\right)^2
+4g^6\cos^2\theta$. At trivial flux, the two cells acquire identical ergotropies,
$W_a(t)=W_b(t)$ [Fig.~\ref{fig2}(a)]. At quarter flux
$\theta=\pm\pi/2$, their charging becomes asymmetric, with
the favored cell switching upon flux reversal. Cell $a$ is
favored at $\theta=+\pi/2$ and cell $b$ at
$\theta=-\pi/2$ [Fig.~\ref{fig2}(b)]. The steady-state
ergotropies follow in closed form by Cramer's rule~\cite{SM},
giving $W_{a(b)}^{\rm ss}(\theta)=\omega_{\rm res} N_{a(b)}(\infty)\propto f^{-1}(\theta)$.
Since $f$ is even in $\theta$ while the cell-dependent terms
are odd, $W_a^{\rm ss}(\theta)=W_b^{\rm ss}(-\theta)$,
and the directional contrast is maximal at quarter flux. The favored cell approaches
$48.9\,\omega_{\rm res}(\Omega_d/g)^2$, whereas the reciprocal cell peaks at $0.676$ near
$gt=2.09$. The ergotropies at $\theta=\pm\pi/2$ vs $gt$ are depicted in Figs.~\ref{fig2}(c) and \ref{fig2}(d), with the reciprocal reference given by the purple curve. The long-time values exceed the reciprocal case by two orders of magnitude, and the insets resolve the early-time oscillations. The average charging powers vs $gt$ at the same two fluxes are depicted in Figs.~\ref{fig2}(e) and \ref{fig2}(f). Nonreciprocal charging raises the charging speed above the reciprocal case.

\begin{figure}
	\centering
	\includegraphics[width=0.42\textwidth]{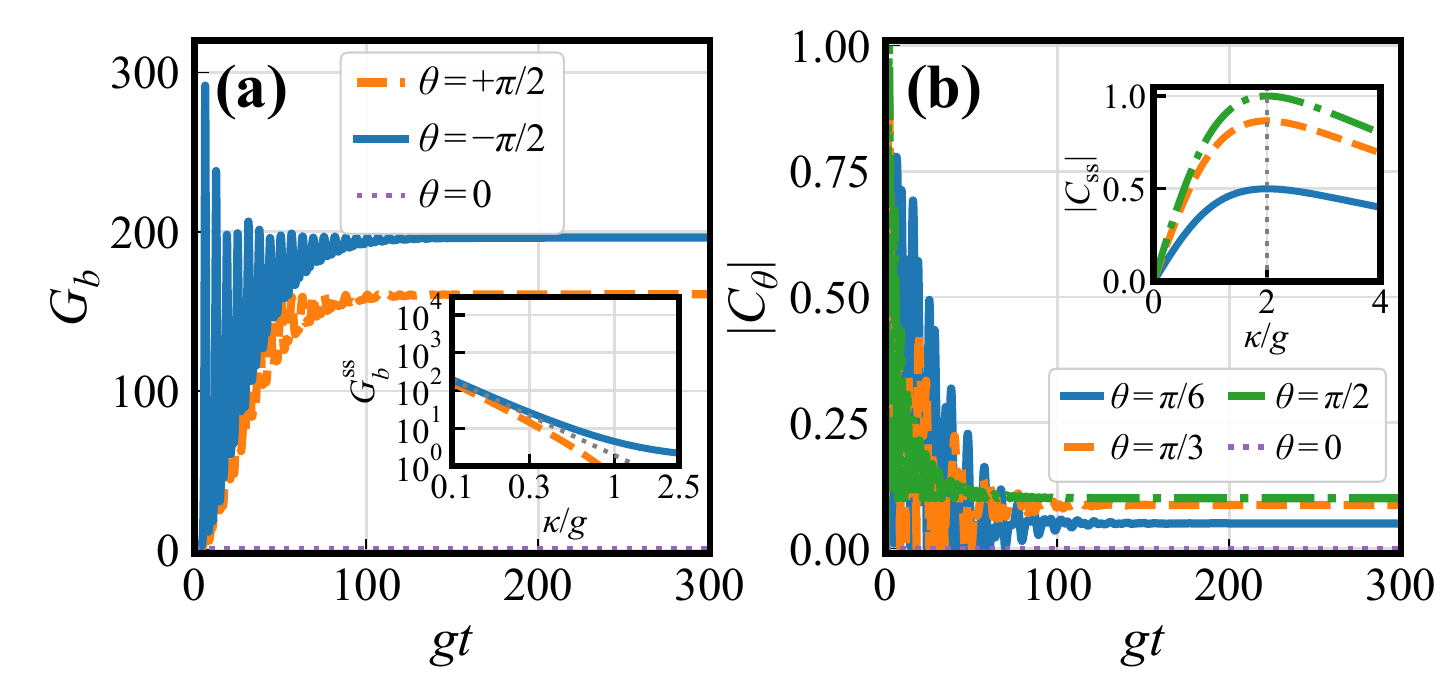}
	\caption{(a) Gain $G_b$ vs $gt$ at $\theta=+\pi/2$ (orange dashed) and $\theta=-\pi/2$ (blue solid), with the dotted curve for the reciprocal reference $\theta=0$ and the inset showing $G_b^{\rm ss}$ vs $\kappa/g$. (b) Contrast $|C_\theta|$ vs $gt$ at $\theta=\pi/6$, $\pi/3$, $\pi/2$, with the dotted curve for the reference $\theta=0$ and the inset showing $|C_{\rm ss}|$ vs $\kappa/g$. Here, $\kappa=0.1g$, $g=\Omega_d=1$, $\Delta=0$.}
	\label{fig3}
\end{figure}

To quantify the advantage of nonreciprocal over reciprocal charging,
we define the dynamical gain
$G_b(t,\theta)=W_b(t,\theta)/W_b(t,0)$, where $W_b(t,0)$ denotes
the reciprocal reference. This gain vs time and flux is depicted in Fig.~\ref{fig3}(a). At the operating point $\kappa=0.1g$,
the gain for $\theta=-\pi/2$ reaches a transient maximum of $291.8$
at $gt=6.27$ before settling to $G_b^{\rm ss}(-\pi/2)=196.3$.
By comparison, the opposite direction, $\theta=+\pi/2$, yields a
lower steady-state gain of $160.7$. This directional asymmetry
motivates a direct comparison between the two flux-reversed
charging protocols. The inset of Fig.~\ref{fig3}(a) shows the steady-state gain
vs $\kappa$. For the reciprocal reference,
the gain saturates as $\kappa\to 0$ because the term
$4g^6\cos^2\theta$ in $f(\theta)$ keeps the denominator finite.
At quarter flux, however, this term vanishes, allowing the favored
ergotropy to grow as $\kappa^{-2}$. Consequently, the gain increases
as the linewidth narrows.

To make this directional asymmetry explicit, we define the
bidirectional contrast ratio
\begin{align}
	C_\theta(t)=
	\left|
	\frac{W_b(t,\theta)-W_b(t,-\theta)}
	{W_b(t,\theta)+W_b(t,-\theta)}
	\right|,
	\label{eqCdyn}
\end{align}
This contrast vs $gt$ is depicted in Fig.~\ref{fig3}(b). In the long-time limit,
$C_\theta(t)$ reduces to $C_{\rm ss}
={g\kappa|\sin\theta|}/{(g^2+\kappa^2/4)}$ the steady-state bidirectional contrast ratio for the uniform ring. Dynamically, the contrast reaches unity. At $gt^*=1.92$, the
cell charged at $\theta=+\pi/2$ is empty, while its flux-reversed
partner has already stored $1.60$, approximately $3.3\%$ of its
steady-state value. The perfect contrast is transient. The two
protocols contain opposite interference terms, $g\kappa\sin\theta$,
and both relax to finite steady states. Accordingly, at $\kappa=0.1g$,
the contrast settles to approximately $0.1$, as shown in the inset
of Fig.~\ref{fig3}(b). The steady-state contrast increases with
linewidth and reaches unity at $\kappa=2g$, where the uniform ring
has a reverse null. Perfect steady-state directional charging
therefore requires both this linewidth condition and nontrivial flux.

{\it Backflow-suppressed charging}.---Having established the steady-state response, we now examine transient backflow. We consider the net energy current into cell $b$. From the
continuity equation for
$W_b=\omega_{\rm res} n_b$ with $n_b=\langle b^\dagger b\rangle$, we obtain $\dot W_b=J_{c\rightarrow b}+J_{a\rightarrow b}-\kappa_b W_b$,
where $J_{c\rightarrow b}
=-2\omega_{\rm res} g_{bc}\Im[\langle c\rangle^*\langle b\rangle]$ and $J_{a\rightarrow b}
=-2\omega_{\rm res} g_{ab}\Im[e^{i\theta}\langle a\rangle^*\langle b\rangle]$. Together, they give the total net energy current from the charger $c$ and cell $a$ into cell $b$, $J_b=2\omega_{\rm res}\,\Im[\langle b\rangle^*S]$,
with the net driving amplitude $S(t)=g_{bc}\langle c\rangle+g_{ab}e^{-i\theta}\langle a\rangle$ acting on cell $b$.
At quarter flux, the dynamics admits an invariant real subspace~\cite{SM}.
Specifically, by setting $\langle c\rangle=x_c$, $\langle a\rangle=ix_a$, and $\langle b\rangle=ix_b$, where $x_c$, $x_a$, and $x_b$ are real, the first-moment equations at resonance preserve this invariant subspace and reduce to
\begin{align}
	\dot{x}_c
	&=-\frac{\kappa_c}{2}x_c
	+g_{ac}x_a+g_{bc}x_b+\Omega_d,
	\nonumber\\
	\dot{x}_a
	&=-g_{ac}x_c-\frac{\kappa_a}{2}x_a
	\pm g_{ab}x_b,
	\nonumber\\
	\dot{x}_b
	&=-S-\frac{\kappa_b}{2}x_b.
	\label{eqmanifold}
\end{align}
The "$\pm$" in the second row corresponds to $\theta=\pm\pi/2$. Thus, the quarter-flux dynamics is confined to an invariant manifold
on which the net driving amplitude $S(t)$ is real. A finite detuning
introduces additional imaginary components and destroys this real
invariant subspace~\cite{SM}. The invariance can also be understood
from an antiunitary symmetry. At quarter flux, the transformed matrix
$\mathbf T\mathbf A\mathbf T^{-1}$, with
$\mathbf T=\mathrm{diag}(1,-i,-i)$, is real. Therefore, if the
transformed variables are initially real, they remain real
throughout the time evolution. This symmetry is absent at
$\theta=0$.

\begin{figure}
	\centering
	\includegraphics[width=0.42\textwidth]{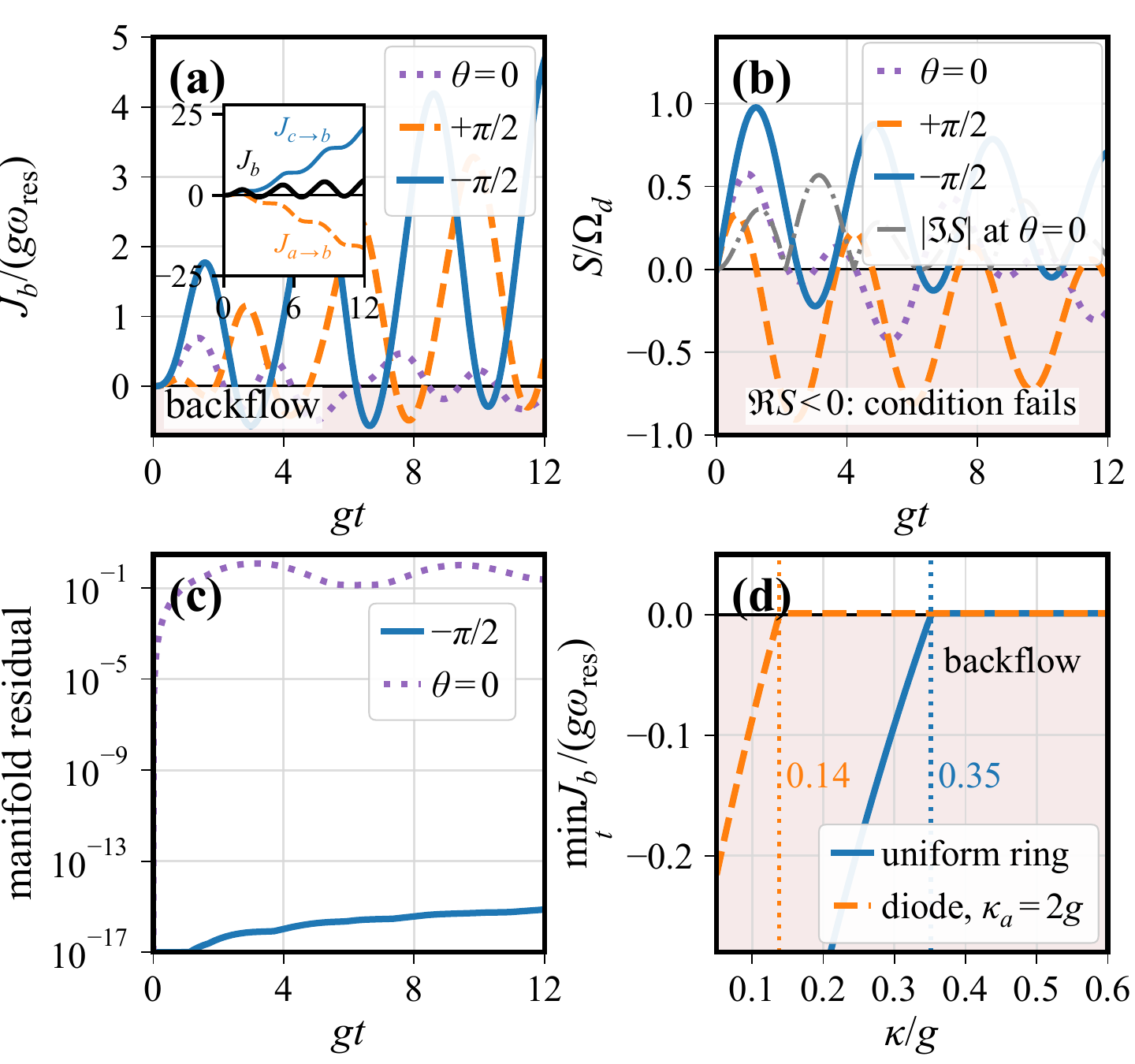}
	\caption{(a) Net current $J_b$ vs $gt$ at $\theta=0$ (dotted), $+\pi/2$ (dashed), $-\pi/2$ (solid); shading marks backflow and the inset resolves $J_{c\to b}$ and $J_{a\to b}$ at $\theta=-\pi/2$. (b) Net driving amplitude $S$ vs $gt$ at quarter flux, gray curve for $|\Im S|$ at $\theta=0$ and shading for $\Re S<0$ failure. (c) Manifold residual $\varepsilon=\max(|\Im\langle c\rangle|,|\Re\langle a\rangle|,|\Re\langle b\rangle|)$ vs $gt$, with the exact result at $-\pi/2$ at round-off level. (d) Minimum net current $\min_t J_b$ vs $\kappa/g$ for the uniform ring and the diode point $\kappa_a=2g$, with lines at $0.35g$ and $0.14g$ and the operating point $0.1g$. Here, $g=\Omega_d=1$, $\Delta=0$.}
	\label{fig4}
\end{figure}

The invariant manifold fixes the phase structure of $S(t)$ but not its sign. The equation for $x_b$ can be integrated exactly, as $x_b(t)=
-e^{-\kappa_b t/2}
\int_0^t e^{\kappa_b t'/2}S(t')dt'$, where we have used the vacuum initial condition $x_b(0)=0$.
If $S(t)\geq0$ throughout the evolution, this expression ensures
$x_b(t)\leq0$. The total current then satisfies $J_b=-2\omega_{\rm res} x_bS\geq0$.
For the vacuum initial state, $S(t)\geq0$ throughout the charging process is a sufficient condition for a nonnegative net current into the storage cell. The current decomposition and the linewidth scans are given~\cite{SM}. The condition constrains the total current rather than each
individual channel. Indeed, on the invariant manifold,
$J_{a\rightarrow b}=2\omega_{\rm res} g_{ab}x_ax_b$, whose sign is not fixed. The total current $J_b$ for the uniform ring at $\kappa=0.1g$ is depicted in Fig.~\ref{fig4}(a). Although the dynamics
lies on the invariant manifold at quarter flux, the current is not
strictly nonnegative. Its minimum is
$\min J_b=-0.569\,g\omega_{\rm res}$ over $0\leq gt\leq12$. With $\kappa_a=2g$, the negative excursion is reduced to
$-0.087\,g\omega_{\rm res}$. The individual channel currents resolved in the inset of Fig.~\ref{fig4}(a) show why the total current can still exhibit backflow. With $\kappa_a=2g$,
$J_{a\rightarrow b}$ is negative in the steady state but positive
during the early transient. The channel currents can therefore
have opposite signs;
their sum, rather than either channel separately, determines
whether energy flows back from the battery.

The net driving amplitude, the manifold residual, and the minimum net current are shown in Figs.~\ref{fig4}(b)-\ref{fig4}(d) to identify the origin of the residual backflow and its linewidth thresholds. As depicted in Fig.~\ref{fig4}(b), the
net driving amplitude changes sign even though the invariant
manifold remains intact. By contrast, at $\theta=0$ the real
invariant manifold is broken, as indicated by the nonzero residual
in Fig.~\ref{fig4}(c). The uniform ring satisfies $S(t)\geq0$ throughout the evolution for $\kappa\gtrsim0.35g$, whereas at $\kappa_a=2g$ the threshold drops to $\kappa_{b(c)}\gtrsim0.14g$ [Fig.~\ref{fig4}(d)]. Concentrating the loss in cell $a$ therefore lowers the linewidth required to remove transient backflow.

\begin{figure}
	\centering
	\includegraphics[width=0.42\textwidth]{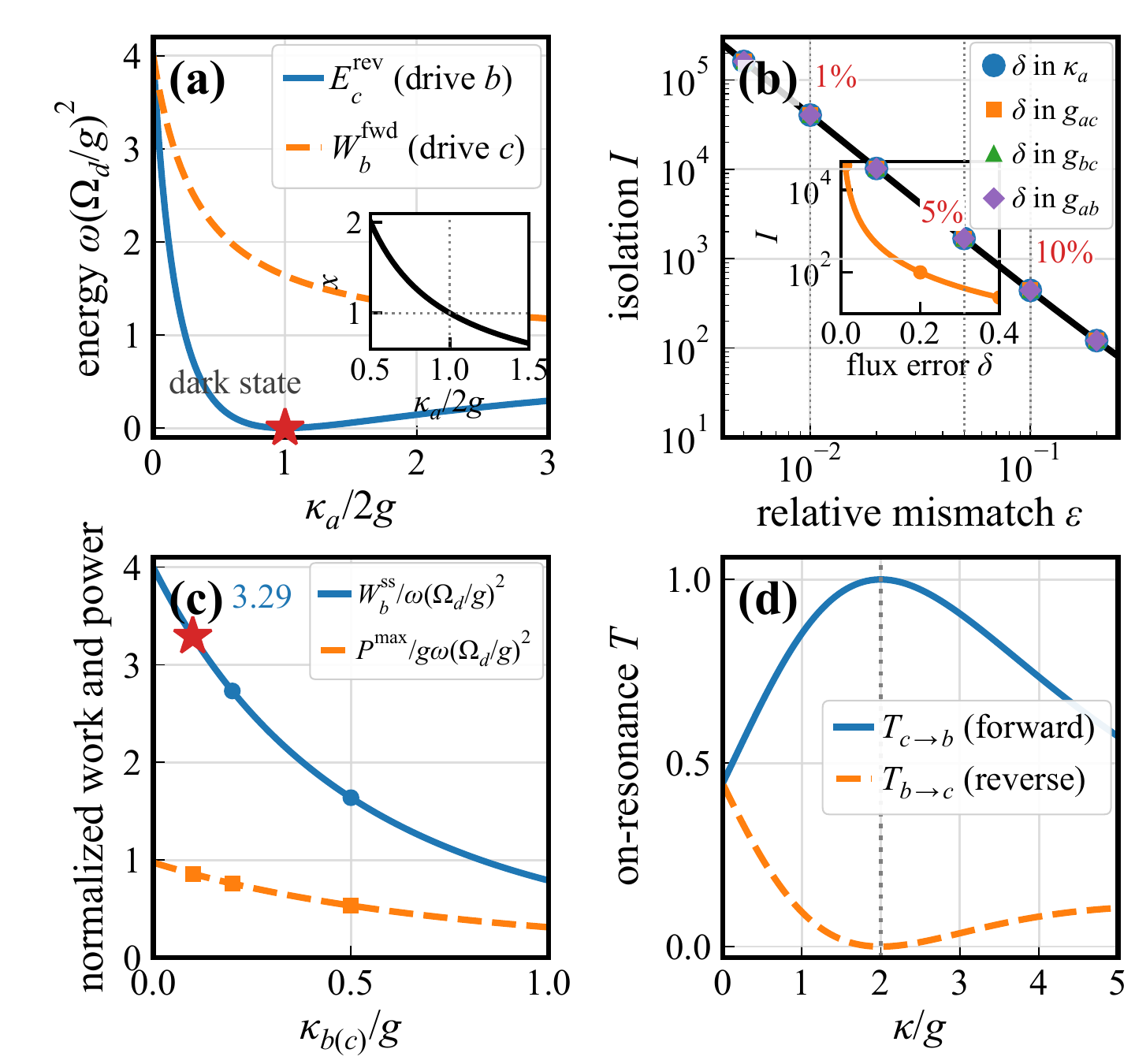}
	\caption{(a) Reverse charger energy $E_c^{\rm rev}$ (drive $b$) and forward battery energy $W_b^{\rm fwd}$ (drive $c$) vs $\kappa_a/2g$, star marking the dark state and the inset its neighborhood. (b) Isolation $I$ vs mismatch $\varepsilon$, with markers for detuning $\delta$ in $\kappa_a$, $g_{ac}$, $g_{bc}$, $g_{ab}$, the black line for the analytical prediction, and the inset for $I$ vs flux error $\delta$. (c) Stored work $W_b^{\rm ss}$ and peak power $P^{\max}$ vs $\kappa_{b(c)}$ at the diode point, star at $3.29$. (d) On-resonance transmission $T_{c\to b}$ (forward) and $T_{b\to c}$ (reverse) vs $\kappa$, the gray dotted line at $\kappa=2g$ marking perfect isolation. Here, $g=\Omega_d=1$, $\Delta=0$.}
	\label{fig5}
\end{figure}

{\it Quantum battery diode.}---\label{sec:diode-main}
We now identify the operating point that yields exact steady-state isolation and call it the diode point. Requiring the charger to remain dark under reverse driving fixes the matching condition. Its physical origin and the resulting isolation are examined below. To derive the condition, we inject energy into cell
$b$ rather than into the charger and require the charger to remain
dark in the steady state, $\langle c\rangle=0$. Setting
$\dot{\mathbf M}=0$, the first row of the steady-state equations
gives $g_{ac}\langle a\rangle+g_{bc}\langle b\rangle=0$. Substituting this relation into
the second row yields $\kappa_a g_{bc}
=2ig_{ac}g_{ab}e^{i\theta}$. For positive coupling strengths, the two sides coincide
only at $\theta=-\pi/2$, giving rise to diode point $\kappa_a={2g_{ac}g_{ab}}/{g_{bc}}$~\cite{SM}. At this location, the reverse-driven ring admits an exact
steady state with $\langle c\rangle=0$, and the two battery amplitudes satisfy $\langle a\rangle/\langle b\rangle=-g_{bc}/g_{ac}$. Consequently,
$W_a^{\rm rev}=(g_{bc}/g_{ac})^2W_b^{\rm rev}$, with $W_b^{\rm rev}=\omega_{\rm res}\Omega_d^2/(\kappa_b/2+g_{bc}g_{ab}/g_{ac})^2$ as detailed in the SM.
For equal couplings and $\kappa_{b(c)}=0.5g$, this gives
$W_b^{\rm rev}=0.64\,\omega_{\rm res}(\Omega_d/g)^2$.
Thus, energy injected into the battery is dissipated through the
lossy cell without producing a steady-state response in the
charger. Under forward charging, the same matching condition
eliminates the reverse transfer from the battery to the charger,
while retaining a finite battery response [Fig.~\ref{fig5}(a)].
The matching condition depends only on the
linewidth of cell $a$ and is independent of $\kappa_b$ and
$\kappa_c$.

Physically, this cancellation is captured by an exact steady-state elimination of cell $a$~\cite{SM}, the most strongly damped mode in the regime considered. This is analogous to the damped-auxiliary-mode elimination used to generate nonreciprocal interactions~\cite{Metelmann2015,Metelmann2017} within the cascaded-input-output framework~\cite{GardinerCollett1985,Carmichael1993}. The two-mode picture explains the mechanism, while the transient requires the full three-mode dynamics. This elimination yields an effective two-mode description involving
the charger and cell $b$, with couplings $G_{cb}=g_{bc}-i\Gamma e^{-i\theta}$ and $G_{bc}=g_{bc}-i\Gamma e^{i\theta}$, 
where $\Gamma=2g_{ac}g_{ab}/\kappa_a$. The elimination also
renormalizes the mode linewidths, $\kappa_c^{\rm eff}
=\kappa_c+{4g_{ac}^2}/{\kappa_a}$ and $\kappa_b^{\rm eff}
=\kappa_b+{4g_{ab}^2}/{\kappa_a}$. At $\theta=-\pi/2$, the effective couplings become
$G_{cb}=g_{bc}+\Gamma$ and $G_{bc}=g_{bc}-\Gamma$.
The matching condition is equivalent to $\Gamma=g_{bc}$,
for which the reverse coupling vanishes while the forward
coupling is doubled. For equal couplings, this gives
$\kappa_a=2g$ (see also the derivation~\cite{SM}). Dissipation localized in 1 battery cell and flux-controlled interference together produce this effective one-way coupling. For a step drive applied to cell $b$ at the diode point~\cite{SM},
{the instantaneous ratio $W_b(t)/E_c(t)$} peaks at $7.1$, $6.2$, and $5.8$~dB for
$\kappa_{b(c)}=0.5g$, $0.2g$, and $0.1g$, respectively, and then
approaches the steady-state null.

The robustness of the diode can be quantified by the ratio of
the forward battery energy to the reverse charger energy. Defining
$x\equiv2g_{ac}g_{ab}/(\kappa_a g_{bc})$, the exact steady-state solution gives
\begin{align}
	I=\frac{W_b^{\rm fwd}}{E_c^{\rm rev}}
	=\left(\frac{x+1}{x-1}\right)^2.
	\label{iso}
\end{align}
For a relative mismatch $\epsilon$ in any one of the matching
parameters, this expression reduces to
$I=(2/\epsilon\pm1)^2$. A flux deviation $\delta$ from the
diode point gives $I=\cot^2(\delta/2)$. These relations are
independent of $\kappa_b$ and $\kappa_c$. The isolation reaches
$46$~dB for a $1\%$ mismatch in one matching parameter,
$32$~dB for a $5\%$ mismatch, and
$26$~dB for a $10\%$ mismatch.
The numerical results for individual
parameter mismatches agree with Eq.~\eqref{iso} and the detailed
error budget~\cite{SM}, while the flux
dependence follows the corresponding cotangent law
[Fig.~\ref{fig5}(b)].

{\it High-$Q$ storage, power, and efficiency.}---
Because the matching condition places the dissipation required for directionality on cell $a$, the storage cell $b$ retains its bare quality factor and is affected only by the coupling-induced linewidth. This separation
allows the storage performance to be improved by reducing
$\kappa_b$ without changing the exact matching condition given above. For $\kappa_{b(c)}$, the stored work rises
from $1.64\,\omega_{\rm res}(\Omega_d/g)^2$ at $\kappa_b=0.5g$ to
$2.73\,\omega_{\rm res}(\Omega_d/g)^2$ at $\kappa_b=0.2g$ and
$3.29\,\omega_{\rm res}(\Omega_d/g)^2$ at $\kappa_b=0.1g$, approaching
$4\,\omega_{\rm res}(\Omega_d/g)^2$ in the joint low-linewidth limit
[Fig.~\ref{fig5}(c)].

The storage advantage is also evident when comparing the diode
with the impedance-matched uniform ring at equal unit contrast.
The latter stores $0.25\,\omega_{\rm res}(\Omega_d/g)^2$ in cell $b$, whereas
the diode stores 11 times more at $\kappa_b=0.2g$ and 13
times more at $\kappa_b=0.1g$. The uniform ring at $\kappa=0.1g$ stores $48.9\,\omega_{\rm res}(\Omega_d/g)^2$ but has no reverse null. At resonance and zero temperature, the diode point gives exact steady-state reverse isolation in the coherent amplitudes at a stored work of $3.29\,\omega_{\rm res}(\Omega_d/g)^2$.

Reducing the storage linewidth also increases the average charging
power. It rises from $0.53$ to $0.76$ and then to
$0.86\,g\omega_{\rm res}(\Omega_d/g)^2$ as $\kappa_{b(c)}$ decreases
from $0.5g$ to $0.2g$ and $0.1g$, respectively. The peak transfer
efficiency follows the same trend, increasing from $0.49$ to $0.65$
and $0.72$. These values exceed the efficiency of the reciprocal
ring at the same storage linewidth, which is $0.23$ at $\kappa=0.1g$~\cite{SM}. The matching condition is also reflected in the
linear-response transmission. The forward transmission approaches
unity, while the reverse transmission vanishes
[Fig.~\ref{fig5}(d)]. 

{\it Experimental realization}.---Toward practical implementation, the flux-threaded bosonic ring can be realized in established superconducting-circuit and cavity-magnonics platforms. In
superconducting circuits, SQUID or SNAIL couplers provide tunable
complex hopping amplitudes, enabling in situ control of the phase
accumulated around the loop~\cite{Blais2021,Roushan2017,
	Chapman2017,Ozawa2019}. For representative coupling strengths
$g_i/2\pi\simeq5$-$30$~MHz, the matching condition requires $\kappa_a/2\pi\sim10$-$60$~MHz,
while the intrinsic linewidths of the charger and storage cell can
remain much smaller, with $\kappa_{b,c}/2\pi\lesssim0.1$-$1$~MHz.
At a frequency of $5$~GHz, this range corresponds to a bare quality
factor $Q\gtrsim5\times10^3$. For example, at the diode point with
$\kappa_b=0.2g$ and $Q_b=5\times10^3$, the ring stores
$2.73\,\omega_{\rm res}(\Omega_d/g)^2$ while exactly satisfying the
steady-state matching condition. At the diode point the net current stays nonnegative for $\kappa_{b(c)}\gtrsim0.14g$ [Fig.~\ref{fig4}(d)], and the example used below lies within this range. For the
representative parameters, the
coupling-induced linewidth renormalization discussed above
reduces the effective storage-mode quality factor to
$Q_b^{\rm eff}\simeq4.5\times10^2$. In situ flux control also
allows the diode response to be reversed on demand. Under these conditions, thermal corrections are negligible, with
$n_{\rm th}\simeq6\times10^{-6}$. Pure dephasing in the lossy cell shifts the matching condition to $\kappa_a=2g-\gamma_{\phi,a}$. Retuning $\kappa_a$ restores the first-moment null, while the resulting mixed state limits the mean-energy isolation to $\simeq1.5\times10^2$ at $\gamma_{\phi,a}=0.05g$~\cite{SM}. At zero temperature and without pure dephasing, the driven modes remain coherent and the stored energy equals the ergotropy~\cite{Allahverdyan2004,SM}. A similar three-mode ring can also be
realized in cavity magnonics, using a YIG Kittel mode coupled to
two cavity modes~\cite{Zhang2014,Tabuchi2014}.

{\it Conclusion.}---
In summary, we have proposed a ring quantum battery comprising a charger and two linked batteries, where a synthetic flux enables nonreciprocal energy transfer and suppresses energy backflow. We demonstrate that backflow can be dynamically suppressed, while at the diode point the reverse coupling is completely switched off and the forward coupling is doubled, enhancing both steady-state storage and peak efficiency. We further investigate the robustness against parameter mismatch, as well as the effects of thermal noise and dephasing, and discuss the experimental feasibility of our proposal. Our work offers a promising route to suppressing energy backflow in practical quantum batteries.

{\it Acknowledgments}.---This work is supported by the National Natural Science Foundation of China (Grant No.~12605044), the Zhejiang Province Key R\&D Program of China (Grant No.~2025C01028), and the Natural Science Foundation of Zhejiang Province (Grant No.~LQN26A040020).

\end{document}